\documentstyle[11pt,paspconf,epsf]{article}

\begin{document}

\title{Young Globular Clusters}
\author{Fran\c cois Schweizer}
\affil{Carnegie Institution of Washington, Dept.\ of Terrestrial Magnetism,
5241 Broad Branch Road, NW, Washington, DC 20015-1305}

\begin{abstract}
Star formation in starbursts appears to be biased toward
compact clusters, with up to 20\% of all stars formed in them.
Observations with {\it HST\,} show that many of these clusters
have luminosities ($-9>M_V>-16$), $U\!BV\!I$ colors,
and half-light radii ($R_{\rm eff}\approx 3$ pc) consistent with
their being young globular clusters (YGC).  Although we know little about
the long-term stability of the youngest clusters ($<$20 Myr), compact
clusters older than $\sim${}$10\, t_{\rm cross}$ (20\,--\,40~Myr) are bound
gravitationally and very likely YGCs.  The present review concentrates on
recent progress achieved in dating such clusters and on evidence that
mergers of spiral galaxies can produce relatively rich subsystems of YGCs
of solar metallicity in the remnants' halos. Studies of such subsystems
suggest a scenario in which second-generation globulars form from giant
molecular clouds squeezed into collapse by the high-pressure
environment of starbursts. These bursts, often driven by mergers, may
explain ongoing cluster formation in NGC 4038/39, the young halo globulars
found around protoellipticals like NGC 3921 and NGC 7252, and the
subpopulations of red metal-rich globulars observed in many giant ellipticals.
\end{abstract}


\section{Introduction}

The study of star clusters in external galaxies has been greatly facilitated
by the advent of the {\it Hubble Space Telescope} and its superb spatial
resolution.  It is now nearly routine to discover hundreds of new
clusters in a galaxy, measure their broad-band colors, and
estimate their ages from these colors under some assumption about their
metallicity.  As this metallicity itself is becoming measurable through
multi-slit spectroscopy on 4-m to 10-m class telescopes, the age-dating
of unresolved star clusters
is bound to increase in accuracy over the next few years.  This review
concentrates on results achieved so far on {\it young} globular clusters
(YGC), by which I mean clusters in the age range $0 \la \tau \la 1$~Gyr.
I also discuss some first results on globular clusters (GC) of intermediate
age ($1\la \tau \la 8$~Gyr) found in elliptical galaxies and on the
possible evolutionary connection between them and the YGCs observed in
galactic mergers.

\section{Galaxies with Young Globular Clusters}

Highly luminous young star clusters, and specifically YGCs, have been
discovered in galaxies of many different types, suggesting that clusters
are a natural byproduct of general star formation (e.g., Elmegreen \&
Efremov 1997).  There seems to be a tendency for massive compact clusters
to form preferentially in the most intense starbursts, with up to 20\% of
the total UV luminosity of such bursts contributed by them (Meurer et al.\
1995; Ho 1997).

Table~\ref{table1} is an attempt to list all galaxies in which YGCs (and
some interme\-diate-age GCs) have been observed in detail.  Among
the 34 galaxies there are 12 ongoing mergers and merger remnants that
have experienced major starbursts, another dozen starburst
galaxies where the bursts' cause remains unknown, and a sprinkle of
galaxies that include four Local-Group members, a blue compact dwarf,
and several barred spirals with circumnuclear rings of star formation.

A crucial issue concerning the youngest star clusters is their exact nature: 
Are they gravitationally bound and will they survive as GCs, or are they
merely compact OB associations that will disperse over the next 10--20~Myr?
The distinction is often difficult to make in starburst and barred galaxies
with ongoing cluster formation, but becomes relatively easy in advanced
mergers and remnants:  Here the mean cluster ages often exceed 40~Myr or
$\sim${}$10\,t_{\rm cross}$, and the case for gravitationally bound YGCs
is strong (\S 3).  Several of the galaxies discussed below are such
advanced mergers.  Mergers have the added advantage that they can be
arranged into some sort of evolutionary sequence, ranging from early
disk--disk mergers like NGC 4038/39 through recent remnants like NGC 3921
and NGC 7252 (Toomre \& Toomre 1972) to suspected old remnants like
NGC 5018 and NGC 3610 (Schweizer \& Seitzer 1992).

Observations of young star clusters in NGC 4038/39, the prototypical pair of
colliding spirals, illustrate the recent progress made with {\it HST}.
Whereas groundbased observations showed some two dozen bright knots of
intense star formation (Rubin et al.\ 1970), images obtained with
{\it HST\,} before refurbishment displayed $\sim$700 young clusters
(Whitmore \& Schweizer 1995), and images obtained since refurbishment now
reveal 14,000 point-like sources (Whitmore et al.\ 1999).  Of these
sources, $>$1000 are likely star clusters, at least 40\% ($\ga$5600 sources)
are likely individual stars, and the remainder are probably a mixture of
individual stars, multiple stars, and poor clusters.  Remarkably, even though
this galaxy pair at 19 Mpc ($H_0=75$) is $\sim$30\% more distant than the
Virgo cluster, three different populations of GCs can now be distinguished
in it: 0--100~Myr old clusters formed during the present collision, a
distinct group of $\sim$500 Myr old GCs presumably formed during a previous
collision, and over a dozen truly old ($\ga$10~Gyr) GCs that must have
belonged to the halos of the input spirals.

Significant populations of YGCs have also been discovered in the two recent
merger remnants NGC 3921 and 7252.  In the former, {\it HST\,} images
obtained with WFPC2 show 102 candidate GCs plus 49 looser associations.
Most detected globulars are young (250--750~Myr), and roughly half occur
in a central region of $\sim$5~kpc radius (Schweizer et al.\ 1996).
In NGC 7252, deep images obtained with WFPC2 now reveal 500 candidate
clusters brighter than $V=26$ (Miller et al.\ 1997). Of these, nearly
3/4 appear to be YGCs ($\tau \approx 400$--600~Myr) of remarkably uniform
color $V\!-\!I \approx 0.65$, about 20\% are likely old GCs ($V\!-\!I
\approx 0.95$), and a few dozen are very young clusters or OB associations
just born in the central molecular-gas disk.

\clearpage
\begin{table}[h]
\caption{Galaxies With Young Globular Clusters.} \label{table1}
\begin{center}\scriptsize
\begin{tabular}{ll}
\tableline
\noalign{\medskip}
Galaxy & References \\
\noalign{\smallskip}
\tableline
\noalign{\medskip}
\noalign{\centerline{MERGERS}}
\noalign{\smallskip}
NGC 4038/39 (Antennae)& Whitmore \& Schweizer 95; Whitmore+ 99          \\
NGC 3256 &              Zepf+ 99                                        \\
NGC 7727 &              Crabtree \& Smecker-Hane 94                     \\
NGC 6052 &              Holtzman+ 96                                    \\
A0035-335 (Cartwheel) & Borne+ 97                                       \\
\noalign{\smallskip}
NGC 3921 &              Schweizer+ 96                                   \\
NGC 7252 &              Schweizer 82; Whitmore+ 93; Miller+ 97; {\it Sp:}
                        Schw.\ \& Seitzer 93, 98                        \\
NGC 1275 &              Holtzman+ 92; Carlson+ 98; {\it Sp:} Zepf+ 95;
                        Brodie+ 98                                      \\
NGC 3597 &              Lutz 91; Holtzman+ 96                           \\
NGC 5018 &              Hilker \& Kissler-Patig 96                      \\
\noalign{\smallskip}
NGC 1316 &              Schweizer 80; Shaya+ 96; Grillmair+ 99          \\
NGC 3610 &              Whitmore+ 97                                    \\
\noalign{\smallskip}
\noalign{\centerline{STARBURSTS}}
\noalign{\smallskip}
NGC  253 &              Watson+ 96                                      \\
NGC 1140 &              Hunter+ 94                                      \\
NGC 1569 &              Arp \& Sandage 85; O'Connell+ 94; De Marchi+ 97;
			{\it Sp:} Ho \&                                 \\
         &              Filippenko 96a                                  \\
NGC 1705 &              Meurer+ 92, 95; O'Connell+ 94; {\it Sp:} Ho \&
                        Filippenko 96b                                  \\
NGC 1808 &              Tacconi-Garman+ 96                              \\
NGC 3034 = M82 &        van den Bergh 71; O'Connell \& Mangano 78;
                        O'Connell+ 95;                                  \\
         &              Gallagher \& Smith 99                           \\
NGC 3310 &              Meurer+ 95                                      \\
NGC 3690 &              Meurer+ 95                                      \\
NGC 3991 &              Meurer+ 95                                      \\
NGC 4670 &              Meurer+ 95                                      \\
NGC 5253 &              van den Bergh 80; Caldwell \& Phillips 89; Meurer+ 95;
			Calzetti+ 97                                    \\
NGC 7552 &              Meurer+ 95                                      \\
\noalign{\medskip}
\noalign{\centerline{OTHER GALAXIES}}
\noalign{\smallskip}
LMC, SMC, M33, M31 &    Most major Local Group galaxies; many references \\
He 2-10	 &              Blue compact dwarf galaxies; e.g., Conti \& Vacca 94 \\
NGC 1019, NGC 1097, &                                                   \\
\ \ NGC 6951, NGC 7469 &
                        Barred galaxies, circumnuclear rings; e.g., Barth+ 95;
                        Ho 97                                           \\
\noalign{\smallskip}
\tableline
\tableline
\end{tabular}
\end{center}
\vskip -0.1cm
\noindent
\baselineskip=6.0pt
{\scriptsize REFERENCES.---Arp, H., \& Sandage, A.\ 1985, \aj, 90, 1163;\ \
Barth, A.J., et al.\ 1995, \aj, 110, 1009;\ \
Borne, K.D., et al.\ 1997, RevMexAA, 6, 141;\ \
Brodie, J.P., et al.\ 1998, \aj, 116, 691;\ \
Caldwell, N., \& Phillips, M.M.\ 1989, \apj, 338, 789;\ \
Calzetti, D., et al.\ 1997, \aj, 114, 1834;\ \
Carlson, M.N., et al.\ 1998, \aj, 115, 1778;\ \
Conti, P.S., \& Vacca, W.D.\ 1994, \apj, 423, L97;\ \
Crabtree, D.R., \& Smecker-Hane, T.\ 1994, \baas, 26, 1499;\ \
De Marchi, G., et al.\ 1997, \apj, 479, L27;\ \
Gallagher, J.S., \& Smith, L.J.\ 1999, \mnras, 304, 540;\ \
Grillmair, C.J., et al.\ 1999, \aj, 117, 167;\ \
Hilker, M., \& Kissler-Patig, M.\ 1996, \aap, 314, 357;\ \
Ho, L.C.\ 1997, RevMexAA, 6, 5;\ \
Ho, L.C., \& Filippenko, A.V.\ 1996a, \apj, 466, L83;\ \
Ho, L.C., \& Filippenko, A.V.\ 1996b, \apj, 472, 600;\ \
Holtzman, J.A., et al.\ 1992, \aj, 103, 691;\ \
Holtzman, J.A., et al.\ 1996, \aj, 112, 416;\ \
Hunter, D.A., et al.\ 1994, \aj, 108, 84;\ \
Lutz, D.\ 1991, \aap, 245, 31;\ \
Meurer, G.R., et al.\ 1992, \aj, 103, 60;\ \
Meurer, G.R., et al.\ 1995, \aj, 110, 2665;\ \
Miller, B.W., et al.\ 1997, \aj, 114, 2381;\ \
O'Connell, R.W., et al.\ 1994, \apj, 433, 65;\ \
O'Connell, R.W., et al.\ 1995, \apj, 446, L1;\ \
O'Connell, R.W., \& Mangano, J.J.\ 1978, \apj, 221, 620;\ \
Schweizer, F.\ 1980, \apj, 237, 303;\ \
Schweizer, F.\ 1982, \apj, 252, 455;\ \
Schweizer, F., et al.\ 1996, \aj, 112, 1839;\ \
Schweizer, F., \& Seitzer, P.\ 1993, \apj, 417, L29;\ \
Schweizer, F., \& Seitzer, P.\ 1998, \aj, 116, 2206;\ \
Shaya, E.J., et al.\ 1996, \aj, 111, 2212;\ \
Tacconi-Garman, L.E., et al.\ 1996, \aj, 112, 918;\ \
van den Bergh, S.\ 1971, \aap, 12, 474;\ \
van den Bergh, S.\ 1980, \pasp, 92, 122;\ \
Watson, A., et al.\ 1996, \aj, 112, 534;\ \
Whitmore, B.C., et al.\ 1993, \aj, 106, 1354;\ \
Whitmore, B.C., et al.\ 1997, \aj, 114, 1797;\ \
Whitmore, B.C., et al.\ 1999, \aj, in press;\ \
Whitmore, B.C., \& Schweizer, F.\ 1995, \aj, 109, 960;\ \
Zepf, S.E., et al.\ 1995, \apj, 445, L19;\ \
Zepf, S.E., et al.\ 1999, \aj, in press.
}
\end{table}

Among other colliding and merging galaxies with YGCs listed in
Table~\ref{table1} are the ``Cartwheel'' ring galaxy and the well-known
peculiar cD galaxy NGC 1275 (see refs.\ in table, where ``{\it Sp:}''
indicates spectroscopic studies of YGCs).

\section{The Nature of Clusters: Young Globulars}

There is much evidence that the majority of bright clusters in ongoing
mergers and recent remnants are young, and that many probably have masses
similar to those of classical old GCs.  Their youth is indicated by blue
colors, high luminosities, and sometimes their location in H\,II regions.
Masses can be estimated from the measured color indices and luminosities
via comparisons with model star clusters and under the assumption that the
stellar IMF in the clusters is normal.  For an assumed Salpeter IMF,
estimated masses typically lie in the range 10$^5$\,--\,10$^7 M_{\odot}$.
For two 10\,--\,20~Myr old clusters in starburst galaxies, dynamical masses
have been estimated directly from measured velocity dispersions and are
$8\times 10^4 M_{\odot}$ and $3.3\times 10^5 M_{\odot}$, respectively
(Ho \& Filippenko 1996), in good agreement with the median mass of
$1.5\times 10^5 M_{\odot}$ for Milky Way GCs.

However, to demonstrate that a young star cluster is a globular one needs
to know both its size and age.  As mentioned above, only if the cluster
is $\ga${}$10\,t_{\rm cross}$ ($\approx$ 20\,--\,40~Myr) old and
still as compact as an old GC can one conclude that it is gravitationally
bound.  There is now strong evidence that most bright, 40\,--\,1000~Myr old
clusters in merger galaxies have half-light radii comparable to those of
Milky Way GCs (median $R_{\rm eff}=3$ pc).  For clusters in the most distant
studied mergers (NGC 1275, 3921, and 7252), only upper limits can be placed
on the median $R_{\rm eff}$, but since the repair of {\it HST\,} even these
upper limits have been reduced to 4\,--\,6~pc.  In nine nearby starburst
galaxies the median radii of clusters measured with the Faint Object Camera
in the UV, where {\it HST\,}'s  resolution is best, are $R_{\rm eff}\approx
3$ pc (Meurer et al.\ 1995). Hence, the young clusters in these merger and
starburst galaxies have similar sizes as Milky Way GCs.

For more than half of the ``MERGERS'' listed in Table~\ref{table1} (beginning
with NGC 3921), the broad-band colors of star clusters measured with
{\it HST\,} yield ages of typically at least a few 100~Myr, corresponding to
$\ga$10$^2$ cluster-core crossing times.  For 12 clusters, these ages have
been verified by spectroscopic observations (\S 4).  Thus, in connection
with their small measured $R_{\rm eff}$ it seems likely that most of
the observed clusters are true YGCs (see refs.\ of Table~\ref{table1}).

For the ongoing ``MERGERS'' of Table~\ref{table1} (first five objects), the
situation
is less clear.  Not all their clusters are as compact as GCs, and of those
that are many are too young for us to ascertain their future cohesion.
Yet, some clusters appear so enormously massive and compact that, despite
their extreme youth ($\la$20~Myr), they seem good prospects for being GCs.
A good case is Knot~S in NGC 4038, imaged with WFPC2 on the PC chip
(Whitmore et al.\ 1999).  This cluster is highly luminous ($M_V=-16$),
7~Myr old, and has a power-law envelope that shows hundreds of individual
stars and extends to $R\ga 450$~pc, making it a super star cluster of some
sort.  Yet, its core is very compact.  The same is true to a lesser extent
for its companion Cluster \#430, while the $\sim$500~Myr old Cluster \#225
shows both a softer core and a distinct radial cutoff to its envelope.
These three clusters are all likely YGCs and may form an evolutionary
sequence that illustrates the ongoing erosion of cluster envelopes through
tidal forces.

\section{Age Dating Young and Intermediate-Age Globulars}

\begin{figure}[t]
\plottwo{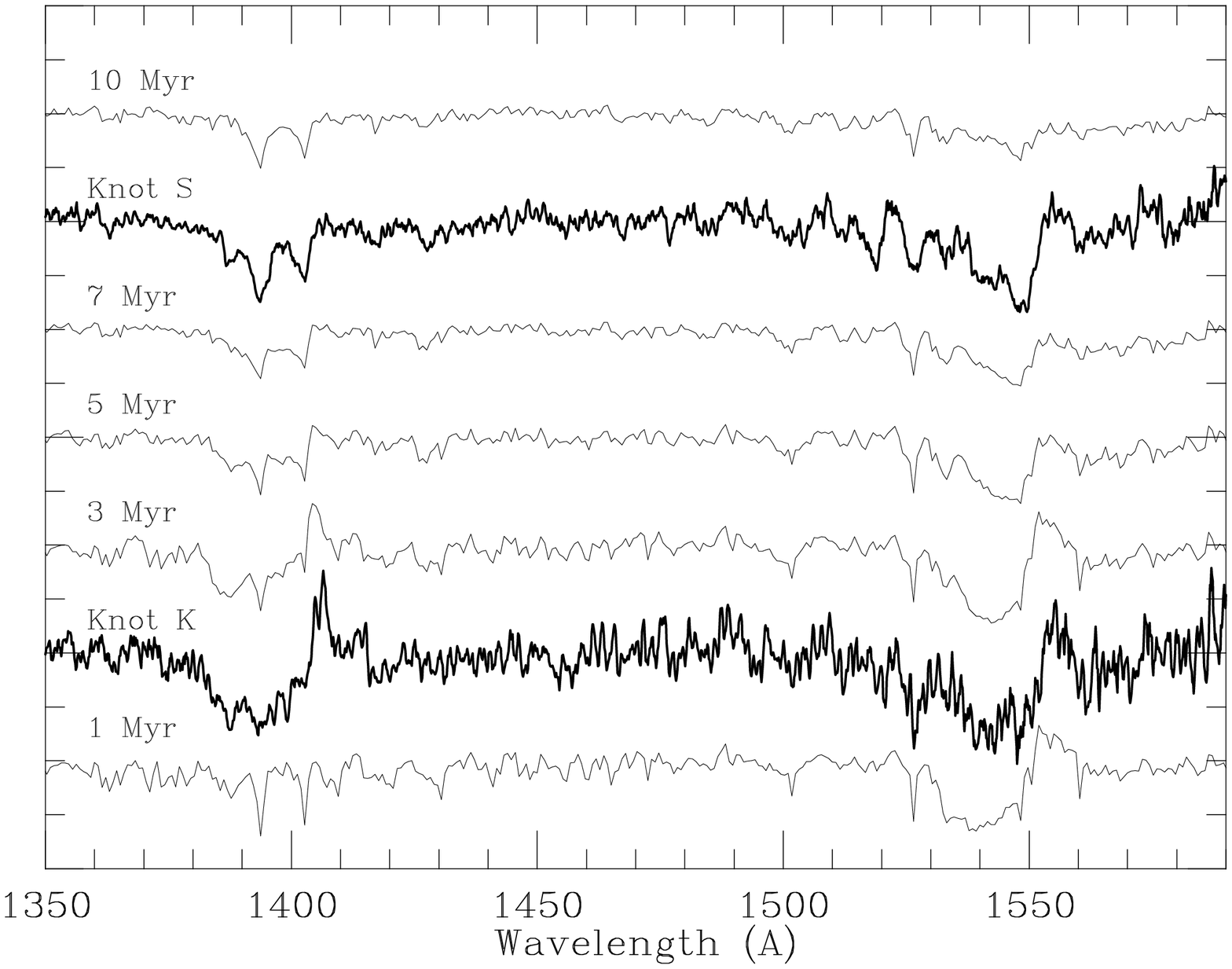}{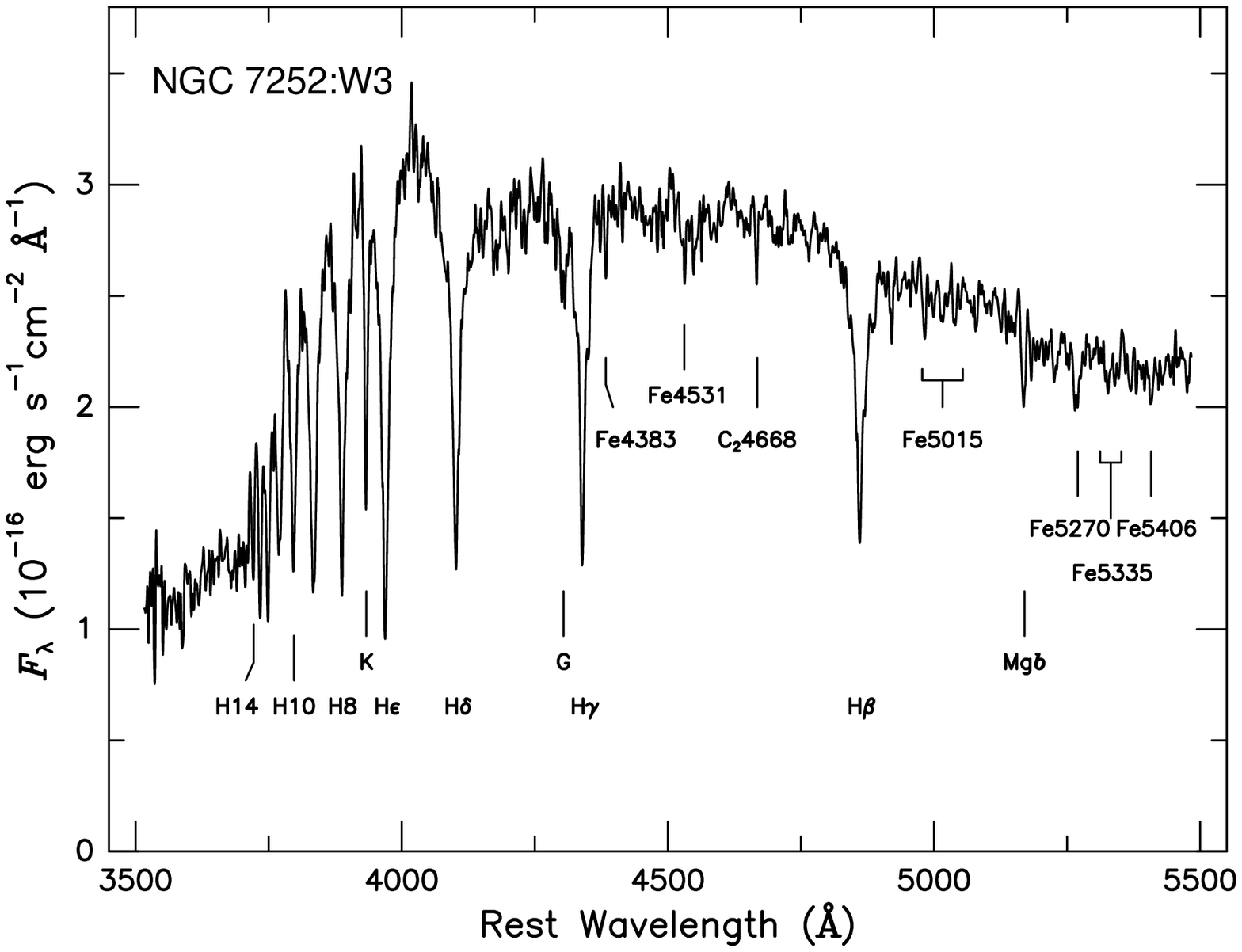}
\caption{Cluster spectra used for age dating: (a) Knots~K and S in
NGC 4038 compared to model-cluster spectra (Whitmore et al.\ 1999);
(b) Cluster W3 in NGC 7252 (Schweizer \& Seitzer 1998).}
\label{fig1}
\end{figure}

Although approximate ages can be estimated for hundreds of clusters from
broad-band (e.g., $U\!BV\!I$) images of the host galaxies, significantly
more accurate ages can be obtained when individual cluster spectra are
available.  Figure~\ref{fig1} shows UV spectra of two $<$10~Myr old
clusters in NGC 4038 and a UV--visual spectrum of the $540\pm 30$ Myr old
cluster NGC\,7252:W3.

For extremely young clusters like Knots K and S in NGC 4038, the space
ultraviolet is the wavelength region of choice for age dating.  The UV
spectra of these two clusters (Fig.~\ref{fig1}a, {\it thick lines}),
obtained with {\it HST\,} and the Goddard High-Resolution Spectrograph
(GHRS), feature strong
stellar-wind lines of Si~IV at $\lambda$1400 and C~IV at $\lambda$1550.
As the comparison with model cluster spectra ({\it thin lines}) shows,
the P Cygni-type profiles of these lines in Knot~K yield a cluster age
of $3\pm 1$ Myr, while the pure absorption-line profiles in Knot~S yield
an age of $7\pm 1$ Myr (Whitmore et al.\ 1999). Both clusters have
UV luminosities of $L_{1500} = 4.4\times 10^{38}$ erg s$^{-1}$ \AA$^{-1}$,
which is an order of magnitude higher than the $L_{1500}$ of R136 in
30 Dor, but still an order of magnitude lower than that of the most
luminous clusters known in extreme starburst galaxies.

For young clusters in the 30\,--\,1000 Myr age range, the strong Balmer
absorption lines from stars that dominate the main-sequence turnoff are a
sensitive, though double-valued age indicator.  Hence, additional lines
like, e.g., the Ca\,II K~line are needed to resolve the age ambiguity.
Figure~\ref{fig1}b shows a spectrum of Cluster NGC\,7252:W3 featuring the
prominent K and Balmer lines plus various Fe and Mg lines.
Figure~2
illustrates the derivation of ages from the Balmer and K
lines for seven YGCs in NGC 7252.  The curves illustrate the evolution
of line ratios and equivalent widths with age for model clusters of
solar metallicity $Z_{\odot}$ (Bruzual \& Charlot 1996 [BC96]), while
the horizontal lines mark values measured for the YGCs. For clusters W3,
W6, and W30, the ages derived from the line ratio K/(H$\epsilon$+H8) are
nearly identical and agree with the higher of the two possible Balmer-line
ages.  If the same is true for the clusters without K-line measurements,
then at least six---and perhaps all seven---YGCs formed 600\,-- 400 Myr
ago, shortly after the onset of the merger (Schweizer \& Seitzer 1998).
\begin{figure}[t]    
\parbox[b]{5.4 cm}{\centering \leavevmode
\epsfxsize=0.40\columnwidth \epsfbox{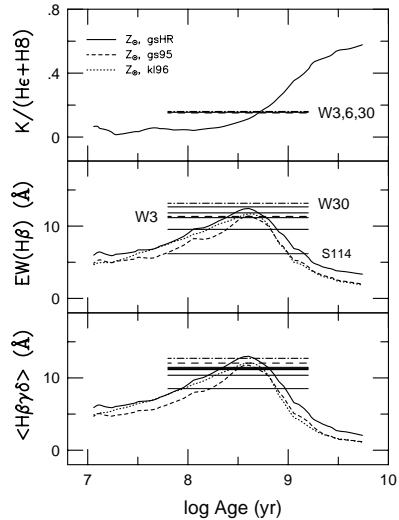}}\hfil
\parbox[b]{8.0 cm}{
\caption[]{Evolution of line ratio K/(H$\epsilon$+H8) and
equivalent widths EW(H$\beta$) and $\langle$H$\beta\gamma\delta\rangle$
in model-cluster spectra with age ({\it curves}, computed from BC96
models), compared with values measured for seven YGCs in NGC 7252
({\it horizontal lines}).  For details, see Schweizer \& Seitzer (1998).
\vspace{7.5mm}}
}
\label{fig2}
\end{figure}

\begin{figure}[t]
\plottwo{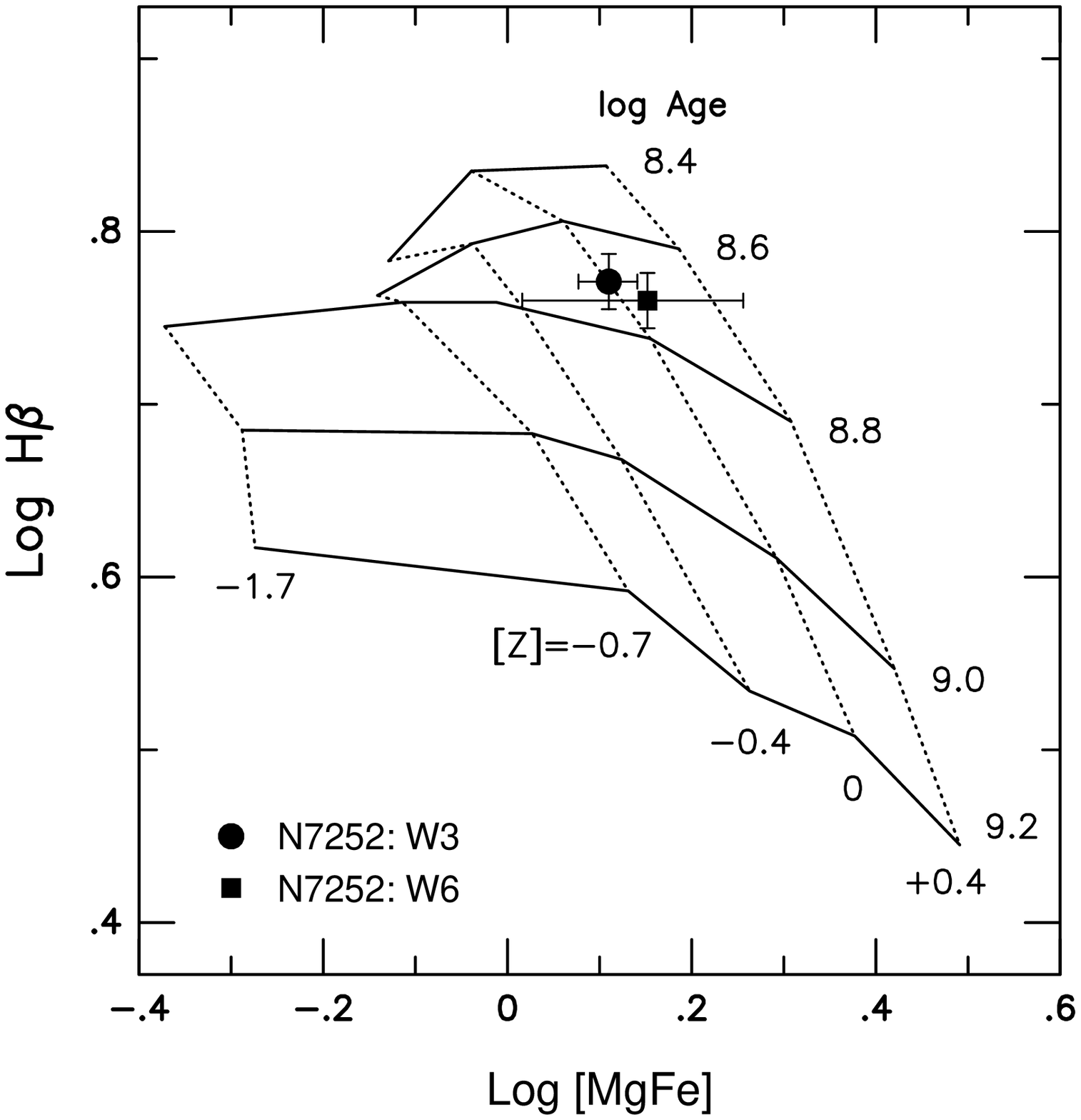}{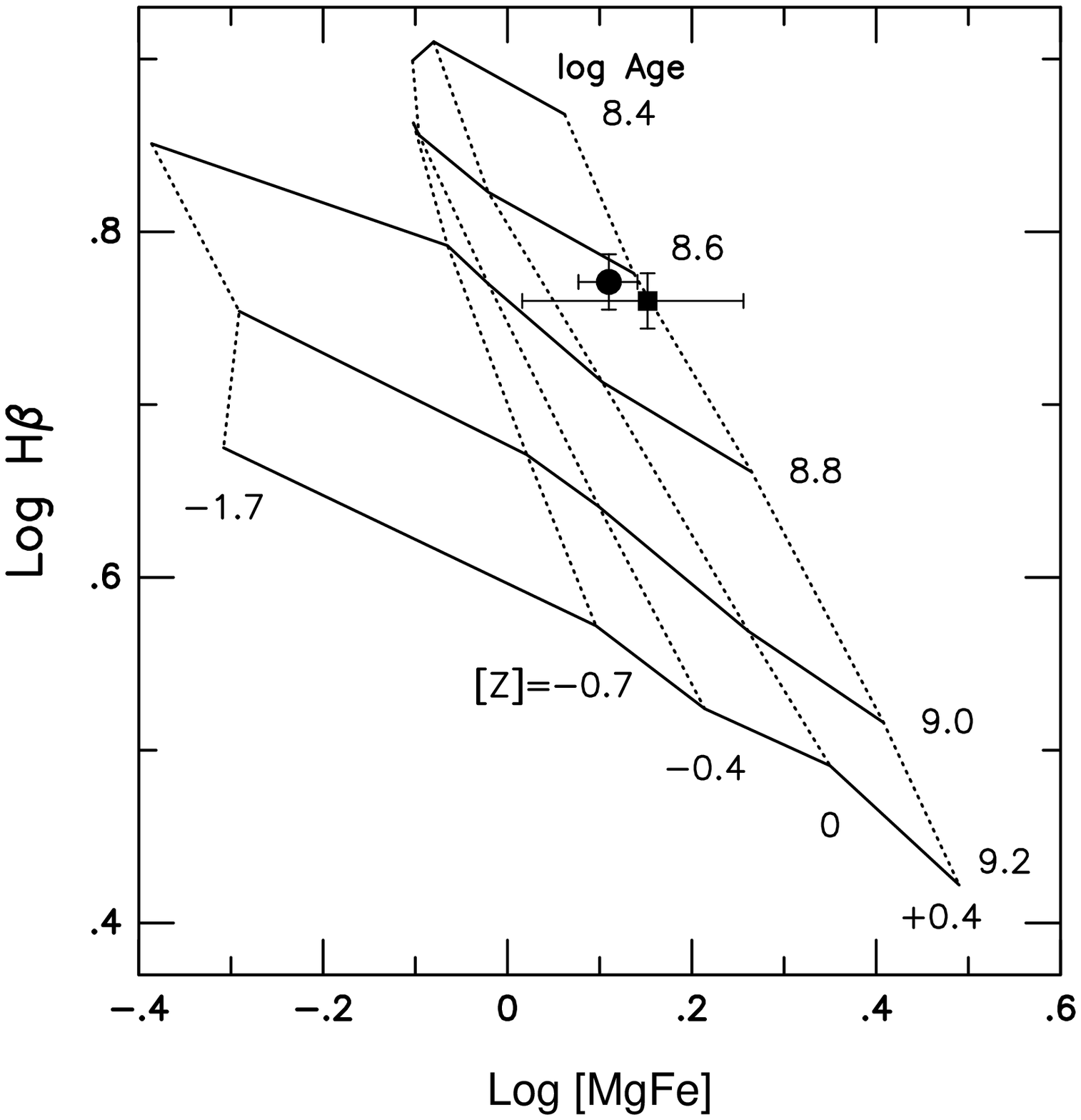}
\caption{H$\beta$--[MgFe] diagrams for two YGCs;
isochrone/isofer grids are based on ({\it left})
Bressan et al.\ (1996) and ({\it right}) BC96 models.}
\label{fig3}
\end{figure}

For the NGC 7252 clusters W3 and W6 the available spectra have
sufficiently high S/N ratios to permit a combined age--metallicity
determination.  Figure~\ref{fig3} shows two versions of the classical
H$\beta$--[MgFe] diagram, where the Lick index H$\beta$ is sensitive
mainly to age and the combined index [MgFe] to metallicity
(Gonz\' alez 1993).  The {\it left} version shows the data points
for the two clusters superposed on a grid of
isochrones ({\it solid} lines) and isofers (lines of constant
metallicity, {\it dotted}) computed from cluster-evolution models by
Bressan et al.\ (1996), while the {\it right} version
shows the same, but for models by BC96.  According to the Bressan et
al.\ models the two clusters have solar metallicity to within
$\pm$0.15 dex,
while according to the BC96 models they have $\sim$2$Z_{\odot}$.
Interestingly, the two model families yield very similar Fe abundances
from individual line-strength indices, but a sharply different Mg
abundance from Mg\,$b$.  It remains unclear whether this difference
reflects model uncertainties or an outright error in one set of
tabulated Mg\,$b$ indices.

This spectroscopic dating confirms previous cluster-age estimates based
on $U\!BV\!I$ colors and leads to the remarkable conclusion that
{\it NGC 7252 possesses a halo population of several hundred YGCs of
solar metallicity}. These clusters have a line-of-sight velocity dispersion
of $140\pm 35$ km s$^{-1}$, comparable to that of the globulars in NGC 5128.
Thus, in NGC 7252 we witness the recent formation of a subsystem of
metal-rich halo GCs.

There are reasons to believe that the metal-rich GCs in elliptical
galaxies with bimodal cluster-color distributions formed during similar,
though more ancient disk--disk mergers (Ashman \& Zepf 1992, 1998;
Schweizer 1987, 1997).  Model simulations show that, as the initially
very blue, second-generation metal-rich globulars age, they become
similar in $V\!-\!I$ color to the old metal-poor GCs at an age of
1\,--\,2 Gyr and then turn distinctly redder. If these simulations
represent reality, we should find ellipticals with intermediate-age
GCs that are slightly redder, but still brighter than the old GCs.
A tentative first example is NGC 3610 (E5), where the slightly
overluminous red globulars indicate an age of $\sim$4 Gyr if of solar
metallicity, or of 6\,--\,7 Gyr if $Z\approx 0.4Z_{\odot}$ (Whitmore et
al.\ 1997).  The former cluster age agrees with the merger age estimated
for NGC 3610 from $U\!BV$ colors (Schweizer \& Seitzer 1992).
Spectra of the red globulars have just recently been obtained with Keck
and LRIS to measure the metallicities and thus help refine the cluster-age
estimates.  Good other candidate ellipticals with intermediate-age
globulars are NGC 5018 and NGC 1316 (see refs.\ in Table~\ref{table1}).

\section{Evolution of GC Subsystems Formed in Mergers}

Intercomparisons of the color distributions and luminosity functions
of GCs in ongoing mergers, merger remnants, and elliptical galaxies
strongly suggest an evolutionary sequence from young to old metal-rich
GC subsystems (Schweizer 1997, esp.\ Figs.\ 1 \& 2; Whitmore 1999,
Figs.\ 2 \& 3).  These subsystems of second-generation clusters appear
to be independent from, and additional to, the well-known systems of old
metal-poor GCs (e.g., Milky Way).

Whereas old metal-poor GCs have many properties that seem to mark them
as first-generation objects formed very early in the Universe's history,
their genesis is understood less well than that of the second-generation
clusters.  Hence, understanding the latter may teach us about the
formation of the former.

There is growing evidence that the second-generation metal-rich clusters
formed from giant molecular clouds (GMC) embedded in the gaseous disks
of spiral galaxies that merged.  First, the mass functions of GMCs and
the luminosity functions of young clusters are power laws with similar
exponents: $N$(GMC)$ \propto M^{-1.6}$ and $N$(GC)$ \propto L^{-1.6\
{\rm to}\ -2}$ (Harris \& Pudritz 1994).  Second, even
the mass ranges of GMCs and globulars are similar:
$\sim$10$^5$--\,$8\times 10^6 M_{\odot}$ and
10$^5$--\,$5\times 10^6 M_{\odot}$, respectively, for Local Group galaxies.
And third, in the merger remnants NGC 3921 and NGC 7252 the YGCs are
distributed radially exactly like the stars ($r^{1/4}$ law), showing that
the progenitors of the YGCs experienced the same violent relaxation as
did the average star (Schweizer et al.\ 1996; Miller et al.\ 1997).  This
implies that the YGCs formed from pre-existing compact progenitors, rather
than from instabilities developing in gas accumulated by the mergers at
the remnants' centers.  The only compact progenitors of sufficient
mass that we know of are the GMCs of the merging spirals.

Interestingly, Jog \& Solomon (1992) predicted that the rapidly
rising pressure of a starburst-heated interstellar medium would
squeeze any embedded GMC into collapse and extremely efficient
($\sim$50\%) star formation, leading to the formation of
massive compact star clusters.  Such a high efficiency seems to
agree with the observed mass ranges of GMCs and globulars
mentioned above.  If so, this merger- and starburst-induced
squeezing is occurring full-scale in NGC 4038/39 and NGC 3256,
has diminished to a trickle in NGC 3921 and NGC 7252, and is
past history in dynamically young ellipticals like NGC 3610.
As we improve the spectrophotometric dating of globular clusters
from their integrated light, we can hope to refine this still
sketchy scenario and perhaps extend it to ellipticals as
old as those formed in clusters like Virgo and Coma.

Support from NSF through Grant AST-95\,29263 is gratefully acknowledged.



\begin{references}

\reference Ashman, K.M., \& Zepf, S.E. 1992, \apj, 384, 50

\reference Ashman, K.M., \& Zepf, S.E. 1998, Globular Cluster Systems
(Cambridge: Cambridge University Press)



\reference Bressan, A., Chiosi, C., \& Tantalo, R. 1996, \aap, 311, 425


\reference Bruzual, A.G., \& Charlot, S. 1996, electronic tables of
cluster models (BC96)



\reference Elmegreen, B.G., \& Efremov, Y.N. 1997, \apj, 480, 235 

\reference Gonz\'alez, J.J. 1993, Ph.\ D.\ thesis, UC Santa Cruz


\reference Harris, W.E., \& Pudritz, R.E. 1994, \apj, 429, 177


\reference Ho, L.C. 1997, RevMexAA, 6, 5

\reference Ho, L.C., \& Filippenko, A.V. 1996, \apj, 466, L83, and
\apj, 472, 600



\reference Jog, C.J., \& Solomon, P.M. 1992, \apj, 387, 152


\reference Meurer, G.R., et al. 1995, \aj, 110, 2665

\reference Miller, B.W., Whitmore, B.C., Schweizer, F., \& Fall, S.M. 1997,
\aj, 114, 2381

\reference Rubin, V.C., Ford, W.K., \& D'Odorico, S. 1970, \apj, 160, 801



\reference Schweizer, F. 1987, in Nearly Normal Galaxies, ed.\ S.M.\ Faber
(New York: Springer), p.~18


\reference Schweizer, F. 1997, in The Nature of Elliptical Galaxies, ed.\
M.\ Arnaboldi, G.S.\ Da Costa, \& P.\ Saha (San Francisco: ASP), p.\ 447
                                 
\reference Schweizer, F., Miller, B.W., Whitmore, B.C., \& Fall, S.M. 1996,
\aj, 112, 1839

\reference Schweizer, F., \& Seitzer, P. 1992, \aj, 104, 1039


\reference Schweizer, F., \& Seitzer, P. 1998, \aj, 116, 2206


\reference Toomre, A., \& Toomre, J. 1972, \apj, 178, 623

\reference Whitmore, B.C. 1999, in Galaxy Interactions at Low and High
Redshift, ed.\ J.E.\ Barnes \& D.B.\ Sanders (Dordrecht: Kluwer), p.~251


\reference Whitmore, B.C., et al. 1999, \aj, in press

\reference Whitmore, B.C., Miller, B.W., Schweizer, F., \& Fall, S.M. 1997,
\aj, 114, 1797

\reference Whitmore, B.C., \& Schweizer, F. 1995, \aj, 109, 960




\end{references}
\end{document}